\documentclass[prd,floatfix,preprintnumbers,letterpaper]{revtex4}
\usepackage{graphicx}
\usepackage{amsmath,amssymb}
\input{epsf}
\usepackage{epsf}
\newcommand {\ga} {\ {\raise-.5ex\hbox{$\buildrel>\over\sim$}}\ }
\newcommand {\la} {\ {\raise-.5ex\hbox{$\buildrel<\over\sim$}}\ } 
\newcommand {\cm} {{\rm cm}}
\newcommand {\GeV} {{\rm GeV}}
\newcommand {\etal} {{\it et al.}}

\begin{document}

\def\be{\begin{equation}}
\def\ee{\end{equation}}

\title{Cosmic Deconstructionism}
\author{Lawrence M. Krauss$^{1,2}$ and Glenn D. Starkman$^{1}$}
\affiliation{$^1$CERCA, Department of Physics, Case Western Reserve University,
Cleveland, OH~~44106}
\affiliation{$^2$Department of Physics \& Astronomy, Vanderbilt University,
Nashville, TN~~37235}
\date{\today}

\begin{abstract}
\bf{Dark Matter that is composed of WIMP remnants of 
incomplete particle-antiparticle annihilation in the early universe 
experiences ongoing annihilation in gravitationally bound large scale structure.
This annihilation will have important consequences in the perhaps distant cosmic future, 
as the annihilation time-scale becomes comparable to the age of the universe.
Much of large scale structure, from galaxy satellites to galaxy clusters will
 disappear.}
\end{abstract}

\maketitle

\section{Introduction}

As demonstrated over 25 years ago by Lee and Weinberg \cite{lee}, and also by
 Dicus and collaborators \cite{tepl}, 
there is a remarkable connection between heavy particles at the electroweak scale 
and the possibility that these particles might be the dark matter that dominates 
galaxies and clusters today.  In particular, the annihilation cross sections 
for these particles can be such that particle-antiparticle annihilations freeze 
out in the early universe at a temperature  somewhat below the mass of the particles. 
Annihilation terminates while still incomplete.  A remnant abundance results 
which, 
depending upon the mass of the particles, can provide a substantial contribution 
to the present mass density of the universe. 

While freeze-out occurs when the universe is a fraction of a second old for these particles, 
annihilations can once again be important at late times.  
For example, as these particles collapse into systems like small satellites of galaxies, 
their density can become sufficiently great so that a small fraction will annihilate, 
producing possible indirect signatures for dark matter which are currently be
ing searched for. 

In this letter, we demonstrate that on time-scales  
much longer than the current age of the universe,  annihilation will once 
again come into equilibrium in the cores of bound structures.  Dark matter
residing in these structures will disappear while matter outside them will
dilute with the expansion of the universe.
We outline here the general expected effects of this annihilation on the
future disappearance of large scale structure.

\section{Eschatological Annihilation Estimates}

The remnant density of Dark Matter determined by freeze-out following annihilations in
the early universe is determined by the canonical freezeout condition:
\begin{equation}
n(T) \langle \sigma v \rangle = H(T)
\end{equation}
where $n(T)$ is the particle number density at temperature $T$, 
$H(T)$ is the expansion rate at this temperature, 
$\sigma$ is the annihilation cross section, 
$v$ is the relative velocity of the particle-antiparticle pair, 
and the average is a thermal average at temperature $T$.   

Since $n \propto T^{3}$, while $H \propto T^2$ at early times, 
for most annihilation cross sections, which themselves fall 
as a positive power of the mean thermal energy at temperature $T$, 
annihilations will fall out of equilibrium in an expanding universe 
as the temperature decreases.

The formation of large scale bound structures with a time-independent density
 profile
adds a new wrinkle to this situation.   
Annihilations become significant on time scales
\begin{equation}
t \ga  t_A \equiv \left(n \langle \sigma v \rangle \right)^{-1}.
\end{equation}
Since $n$ is larger in more tightly bound systems, $t_A$ is smaller there.
As we shall describe in the next section, 
this implies a future history of the universe 
in which structures at increasingly large scales begin to unravel --
a process we term Cosmic Deconstruction.

We first estimate the approximate times for the deconstruction of a variety of cosmic structures --
galaxy satellites, galaxies, and galaxy clusters.
We make these estimates this for realistic SUSY WIMP candidates, 
for which freezeout occurs near the electroweak scale,

For s-wave annihilations, $\sigma v$ is a constant at low energies. 
For particles of mass $m$, annihilating via exchange of a particle of comparable mass $M$, 
$\sigma v \approx \alpha^2 m^2M^{-4}$. 
Here $\alpha$ represents some effective fine-structure constant 
appropriate to the interaction in question.   
Assuming $\alpha \approx 10^{-2}$, and $M \approx 100\GeV$, one finds 
\begin{equation}
\sigma v \approx \left(\frac{m}{M}\right)^2 \times 10^{-26}  \cm^3\sec^{-1}
\end{equation}

Taking as a fiducial estimate the average galactic density 
$\rho_0 \approx 1 \GeV/\cm^3$ (and assuming that $m \approx M \approx100\GeV$),
we find an annihilation time 
\begin{equation}
t_A \approx 10^{11} \frac{n}{n_0}  t_0
 \end{equation}
 where $t_0$ is the present age of the universe.

Clearly, the effect of annihilation on large scale structure is an eschatological issue.

We next estimate what the impact of such annihilations will be on the evolution of structure.  
Not surprisingly, the effects vary depending upon scale.

\section{The Disappearance of Large Scale Structure?}

Large scale structure formation in a $\Lambda$CDM universe is hierarchical, with smaller structures
decoupling from the background distribution when the universe was at a higher
 density.   Numerical 
simulations suggest a general NFW \cite{nfw} form for the dark matter density
inside these structures:
\begin{equation}
\rho(r=xr_s) = \rho_s x^{-1} (1+x)^{-2}
\end{equation}
at the time the halo first decoupled from the cosmic expansion and began to collapse.   
The inner parts of the distribution can be even more strongly clumped with $\rho \approx r^{-1.5}$.   
Ultimately this behavior flattens out in some dense inner core.  

It is predicted that dark matter halos contain many smaller sub-clumps of higher density\cite{HofmannSchwarzStoecker}.   
The number of clumps with mass greater than or equal to $M$ 
relative to the mass  of the host galaxy is \cite{DiemandMooreStadel}
\be
N_{\rm clump}\left(\leq M\right) \approx 10^2 \left(\frac{M_{\rm host}}{M}\right)^2 .
\ee

Fitting to numerical simulations from the galaxy size downward to clumps as small as  $10^{-6}M_{\rm solar}$
one finds central densities that go roughly as $\rho\simeq M^{1/4}$, where $M$ is the mass of the clump.  
Few such clumps are observed in luminous matter, 
in particular, far fewer satellite galaxies are observed than are predicted in N-body simulations. 
This could be because small subclumps are almost entirely dark matter, or because they may not
survive tidal interactions.  
Diemand \etal \cite{DiemandMooreStadel} have argued that at least the cores of such clumps
survive tidal interactions over long times. 

It has not gone unnoticed that dark matter annihilation provides one way of explaining
the absence of these sub-clumps \cite{annihilatingDM}; 
however, $t_A$ is far too long for generic WIMP dark matter to have played any significant role to-date
in the evolution of dark matter clumps.
Moreover, good constraints exist on this possibility 
if the annihilation products include (as they almost always do)
photons or standard model antiparticles (for example \cite{ferr}).)  
We also note that, when it was first claimed that observed density profiles in galactic cores were 
less steep than those predicted in N-body simulations,
it was proposed that enhanced dark-matter particle-antiparticle annihilation 
cross sections 
might smooth out galaxy halo cores (see, e.g., Ref. \cite{Kaplinghat}).    
The cross sections required for this to occur by the present time 
are again far larger than those discussed in the last section.
However, it has been noted that, for sufficiently high densities, 
even the small rate of dark matter annihilations predicted 
for realistic annihilation cross sections of WIMPs might 
provide observable signatures in indirect detection experiments \cite{beac1,beac2,Zentner,zahar,Beacom,ferr}.

We sketch out below the general features that will affect the evolution of structure 
as dark matter annihilation begins to become significant.  
The details will, as we shall describe, depend upon the dark matter to baryon
 content in the system. 

We shall approximate dark matter halos as thermal distributions,
 with particles distributed in velocity space with a roughly 
 Maxwell-Boltzmann distribution $N(x) \approx x^{1/2} \exp{-x}$, 
 where $x =v^2/{v_0}^2$, and $v_0$ is the mean velocity dispersion in the system.

As dark matter begins to annihilate in the dense inner core, there will be three major effects:
\begin{itemize}
\item{ Flattening of Cusps: Dark matter annihilation will maintain the central core density at 
$n_0\approx1/\sigma v t$.  Hence, over time, core densities will decrease.  
We can use this relation to estimate the rate of mass loss from the core (see next item)}
\item{Adiabatic Expansion: Because the timescale for annihilation 
will far exceed the dynamical relaxation time of galactic systems,
the mass loss due to annihilation will result in adiabatic expansion of the core, 
so that  the quantity $M(r_0) r_0$ remains constant as the mass decreases in 
the core \cite{richstone}.   
This adiabatic expansion will supplement annihilation in reducing the density
 in the core.  
As a result, the mass loss due to annihilation will be less than it would otherwise be.   
If we require that the core density fall inversely with time, as given above,
 and use the adiabaticity constraint, then we derive a simple approximate mass loss rate from the expanding core, 
given by $ \dot{M} = M(t)/4t$, instead of $ \dot{M} = M(t)/t$, 
as it would otherwise be. 
Thus, the mass in the core falls as $M=M_0/t^{1/4}$ rather than as $t^{-1}$ 
as it would do otherwise.}
\item{Evaporation:  As mass is lost, the gravitational potential will decrease outside of the core, and particles (both dark matter and baryons) whose velocities were close to the escape velocity before will now escape.  We estimate this effect below, which could be the most significant factor affecting structure deconstruction, depending upon the extent to which dark matter, rather
 than baryons, dominates the potential in the core evaporating region, as we 
describe below} 
\end{itemize}

To get an estimate of the total mass lost due to evaporation for a dark matter dominated system we consider the magnitude shift of the gravitational potential from the time that dark matter begins to annihilate until it completely 
disappears from in the system via a combination of annihilation and evaporation.   
Assuming that particles with velocity squared $v^2 2 {v_0}^2$ 
will initially escape the system, 
if the potential after annihilation is $|V|= p |V_0|$, where $ p< 1$, 
then the fraction of particles that will escape 
will be roughly the fraction in the initial distribution 
with initial velocity-squared exceeding $2 p {v_0}^2$   
This is given by
\begin{equation}
F =  {{\int^2_{2p}{ x^{1/2} e^{-x}d x}} \over {\int^2_{0}{ x^{1/2} e^{-x}d x}} }
\end{equation}

Note that this approximation assumes that non-annihilating particles like baryons do not contribute significantly to the gravitational potential themselves until late times.   
For example, if we consider the dark halos, where the integrated baryon mass 
is perhaps $10 \%$ of the total mass in the core, so that if a significant fraction of the dark 
matter were to annihilate, then  $p=0.1$, and F =.92, and 
most of the baryons and the remaining dark matter will also evaporate.

Once enough dark matter has annihilated in cores, however, so that baryons begin to dominate, the future evolution will change.  The core size will be fixed, determined by the baryon dynamics.  Those baryons in the core that do not
 evaporate during the annihilation process will leave a bound stable remnant.

As annihilation proceeds, the core radius will not grow significantly because
 the core mass will now be dominated by baryons and thus the variation in $M(r)$ will reduced.   We then expect that the outer halo halo will puff up slightly in response to the reduced mass in the core.  Eventually this will stabilize as the dark matter mass in the core continues to reduce, at which point 
annihilation in the outer halo should begin to become significant.  As this process proceeds, from the inside outward, much of the rest of the halo will continue to grow in size, and decrease in density.  
This will result in mass loss due to a combination of annihilation 
and evaporation of the dark matter and baryon particles outside this core, 
as given by eq. 7.  
This halo should then dissipate completely by this combination. 
A significant fraction of the dark matter will evaporate
rather than annihilate, 
and over 90 $\%$ of the halo baryons will evaporate, 
assuming that the dark-matter-to-baryon mass ratio outside the core 
is larger than a factor of 10 .  

In some cases, the residual baryons left 
after the dark matter abundance in the core annihilates down 
will come to dominate the entire potential of the system.  
In this case, the dark matter halo dynamics will be determined 
by the remnant baryonic core, and its distribution will be determined 
by the baryon potential.  Dark matter halos will then continue to annihilate,
with minimal evaporation, because the dominant potential will not be changing
.

Galaxy clusters are simpler to consider.  While the mean density in their inner cores is smaller than for galaxies, and thus annihilation will take longer
 to become significant, the gravitational potential wells in these systems is
 dominated by dark matter throughout.  As a result, after most of the dark matter in the core annihilates, we expect $p<< 1$, so that $F$ will be large.  
In this case, $F$ can roughly be interpreted as the probability for a
cluster galaxy to become unbound.    Assuming that the dark matter continues 
to dominate the potential, the system will continue to puff up in size, with 
core density, and halo density decreasing over time, and with dark matter continuing to evaporate from the system.      

There is, of course, an important caveat in all of our discussions.  The time
scales we are considering are {\it very long}, in excess of perhaps $10^{10}$
 times the current age of the universe.    The extent to which dynamical friction will have caused most of the baryons to have settled in the central region of the galaxy  or galaxy cluster by the point at which annihilation will be significant will dramatically affect whether or not the baryons evaporate following dark matter annihilation and evaporation.  Eventually of course, as 
has been noted in the past, possible black hole evaporation and proton decay 
may drive a much later period of evaporation of any such residual structures 
\cite{adams}.

\section{Conclusions}

Annihilating dark matter, which froze out early in the history of the universe will, because of the formation of bound large scales structures, ultimately
 disappear from large scale structure via annihilation and subsequent evaporation, which become significant in a sufficiently old universe.  We have estimated here that as a result most clusters will disintegrate, and galactic systems will be largely dissipated, although dense baryon cores can remain.   In order to explore in more detail the role of annihilation versus evaporation vs gravitational settling over the incredibly long dynamical timescales we have described here, we plan to carry out numerical simulations    It is also worth pointing out the recent result that regardless of how efficient annihilation is at removing dark matter from large scale structures in the future, nevertheless matter will continue to dominate over the possible relativistic products of this annihilation for all times \cite{kraussscherrer}.

\acknowledgements 
This work was supported in part by the Department of Energy and NASA.  We thank Chris Mihos and Bob Scherrer for several useful discussions.

\end{document}